\documentstyle[sprocl]{article}

\begin{document}

\title{CUBIC INTERACTION VERTICES FOR HIGHER SPIN FIELDS}

\author{R.R. METSAEV}

\address{P.N. Lebedev Physical
Institute,  Leninsky prospect 53, 117924 Moscow, Russia}

\maketitle\abstracts{We present results of investigation into a
problem of cubic interaction vertices of
massless higher spin fields which transform in arbitrary irreps
of the Poincare algebra}

{\it Motivation}.
At present time it is clear that there is no
self-consistent interaction for single higher spin field. It
does not matter whether we take massive or massless fields. It
turn out that while considering higher spin interactions it is
necessary to introduce a infinity tower of all higher spin
fields. Moreover it turns out that besides of minimal
interaction it is necessary to include in game all possible
non-minimal interactions.
For the case of massless higher spin fields a
need for infinite tower of massless fields as well as
non-minimal interactions has explicitly been demonstrated in
Ref.\cite{VAS1}, where a completely
self-consistent equations of motion for anti-de Sitter massless
higher spin fields have been found.  Aforesaid is a reason why
in studies of higher spin fields we are interested in all
possible interactions vertices which can be constructed for
these fields.  Our results generalize those of
Refs.\cite{BER2,BEN2} which were devoted to the analysis of
higher spin interactions in $4D$ space--time.  Note that a full
list of higher spin cubic vertices for totally symmetric
massless fields for any $D$ was obtained in
Refs.\cite{FM1,MET4}.  In this work we solve the problem for
both totally and mixed symmetry fields on a equal basis. We
believe that well known difficulties of flat-space massless
higher spin fields can be overcome by introducing of infinite
tower of fields.

{\it Setting up the problem}.
We start from the  Poincare algebra $ios(D-1,1)$
written as
$$
{} [P^\mu,\,J^{\nu\rho}]={\rm i}\eta^{\mu\nu}P^\rho-\ldots\,,
\qquad
{} [J^{\mu\sigma},\,J^{\nu\rho}]
={\rm i}\eta^{\sigma\nu}J^{\mu\rho}-\ldots
$$
(with $\eta^{\mu\nu}=(+1,-1,\ldots,-1)$),
where $D$ is a space--time dimension. Because we are going to use
light cone  treatment of relativistic dynamics we split
coordinates $x^\mu$ as $x^\mu\rightarrow \{x^\pm,\, x^I\}$,
where $x^\pm=(x^0\pm x^{d+1})/\sqrt{2}$,
$I=1,\ldots, d$ ($d=D-2$) and consider $x^+$ as an evolution
parameter.  In light cone gauge on-mass shell free massless
field is described by tensor field
$\Phi^{I_1\ldots I_n}(p^I,\beta)$
which depends on momenta $p^I$, $\beta \equiv p_-$. The
transversal indices $I_1,\ldots I_n$ come from Young tableaux of
a representation of the SO($d$) group labeled by vector
${\bf m}=(m_1,\ldots, m_\nu)$, $\nu=[d/2]$, which is a weight of
a representation of SO($d$) group.  Let us indicate the
corresponding massless fields as $\Phi_{{\bf m}}(p^I,\beta)$.
Now the problem under consideration can be formulated as
follows: Given three free massless fields
$\Phi_{{\bf m}_a}(p_a^I,\beta_a)$, $a=1,2,3$, construct all
possible tree level local and Poincare invariant interaction
vertices (precise formulation of locality condition is given
below). In what follows we use Fock space vector defined by
$|\Phi\rangle=\Phi^{I_1\ldots I_n}
\alpha_1^{I_1}\ldots \alpha_\nu^{I_n}|0\rangle$, where
$\alpha_l^I$, $l=1,\ldots,\nu$, are the creation operators.
The $|\Phi\rangle$ satisfies the commutation relation
$[|\Phi(p,\beta)\rangle,\,\langle\Phi(p^\prime,\beta^\prime)|]=
\delta(p+p^\prime)\delta(\beta+\beta^\prime)/(2\beta)$.

In order to find Poincare invariant interaction
vertices we construct realization of the Poincare algebra
commutation relations.
In light cone approach the Poincare algebra
splits into a kinematic part spanned by the generators
$G^{kin}=\{$$P_-$,$P^I$,$J^{IJ}$,$J^{+-}$,$J^{+I}\}$,
and dynamic part spanned by $G^{dyn}=\{$$P_+$,$J^{-I}\}$.
The $G^{kin}$ are realized quadratically in the
physical fields while the $G^{dyn}$ are
realized non--linearly. At a quadratical level both $G^{kin}$ and
$G^{dyn}$ have the following representation
$G=\int \beta d\beta d^dp\langle \Phi| r(G) |\Phi\rangle$,
where differential operators $r(G)$ provide realization of
Poincare algebra on the physical fields
$$
\begin{array}{llll}
\!\!r(J^{IJ})=\hat{x}^Ip^J-\hat{x}^Jp^I+M^{IJ}\,,\quad
\!\!r(P_-)=\beta\,,&\!\!r(J^{+I})=-\hat{x}^I\beta\,,&\!\!r(P^I)=p^I,
\\
[5pt]
\!\!r(J^{-I})=\hat{x}^-p^I-\hat{x}^Ih-\frac{1}{\beta}M^{IJ}p^J\,,
&\!\!r(J^{+-})=-\hat{x}^-\beta\,,&\!\!r(P_+)=h,
\end{array}
\eqno{(2)}$$
where $h\equiv p^Ip^I/(2\beta)$,
$\hat{x}^I={\rm i}\partial/\partial p^I$,
$\hat{x}^- =-{\rm i}\partial/\partial \beta$
and
$M^{IJ}$ are the SO$(d)$ group generators.

Now we are looking for cubic corrections to the dynamic
generators $H=P_+$ and $J^{-I}$.
Making use of commutation relations of
$G^{kin}$ with $G^{dyn}$ in cubic approximation
$[G^{kin},\,G^{dyn}]\sim G^{dyn}$
and commutation relations $[G^{dyn},\,G^{dyn}]\!=0$ one can
establish the following representation for $G_{(3)}^{dyn}$
\begin{eqnarray*}
&&
H_{(3)}=\int
d\Gamma_3\,\prod_a\langle\Phi(p_a)|h_3\rangle\,,
\\
&&
J_{(3)}^{-I} =\int
d\Gamma_3\,\biggl(\prod_a\langle
\Phi(p_a)|j_3^{-I}\rangle
-\frac{1}{3} \Bigl(\sum_a\hat{x}_a^I\prod_a
\langle\Phi(p_a)|\Bigr)|h_3\rangle\biggr)\,,
\end{eqnarray*}
\begin{equation}\label{hamden}
\!|h_3\rangle=h_3({\bf P},\beta,M)
|0\rangle_1|0\rangle_2|0\rangle_3\,,
\quad
\!\!\!|j_3^{-I}\rangle=
\frac{2}{3{\bf P}^2}\sum_a
\check{\beta}_a
({\rm i}{\bf P}^I\beta_a\frac{\partial}{\partial \beta_a}
+M_a^{IJ}{\bf P}^J)|h_3\rangle\,,
\end{equation}
where
$$
{\bf P}^I
=\frac{1}{3}\sum_{a=1}^3\check{\beta}_a p_a^I\,,
\qquad
\check{\beta}_a\equiv \beta_{a+1}-\beta_{a+2}\,,
\quad \beta_4\equiv \beta_1\,,
\quad \beta_5\equiv \beta_2
$$
and $d\Gamma_3=\delta(\sum_a p_a^I)\delta(\sum_a\beta_a)\prod
d^dp_ad\beta_a$. The letter $M$ as argument of $|h_3\rangle$
indicates SO$(d)$ group spin degrees of freedom. In addition the
$|h_3\rangle$ should satisfy the equations
\begin{equation}\label{rinvhom}
({\rm i}{\bf P}^J\frac{\partial}{\partial {\bf P}^I}-
{\rm i}{\bf P}^I\frac{\partial}{\partial {\bf P}^J}
+\sum_a M^{IJ}_a)|h_3\rangle=0\,,
\qquad
(\sum_a\beta_a\frac{\partial}{\partial \beta_a}
+{\bf P}^I\frac{\partial}{\partial {\bf P}^I})|h_3\rangle=0\,.
\end{equation}
The 1st equation reflects invariance of the
$|h_3\rangle$ under transversal rotation while the 2nd
equation tells us that $|h_3\rangle$ is a zero-degree
homogeneity function in ${\bf P}^I$ and $\beta_a$.
Besides we impose on the $|h_3\rangle$ the equation
$({\bf P}^I\partial/\partial {\bf P}^I-k)|h_3\rangle=0$ which
tells that the $|h_3\rangle$ is a $k$-degree homogeneity function
in momenta. Formally, the $|h_3\rangle$ constrained by equations
(\ref{rinvhom}) provides a solution to commutation relations
in cubic approximation. To choose physical relevant
$|h_3\rangle$ we impose the following additional condition:
demand that $|h_3\rangle$
and $|j_3^{-I}\rangle$ be monomial in
${\bf P}^I$.  This condition we shall call as locality
condition.  The locality condition amounts to requiring the
$|h_3\rangle$ satisfies
the equation
\begin{equation}\label{loc}
\sum_a\check{\beta}_a ({\rm i}{\bf
P}^I\beta_a\frac{\partial}{\partial \beta_a} +M_a^{IJ}{\bf
P}^J)|h_3\rangle={\bf P}^2|R^I\rangle\,,
\end{equation}
where
$|R^I\rangle$ is a monomial in ${\bf P}^I$.  The equations
(\ref{rinvhom},{\ref{loc}) constitute a complete system of
equations on $|h_3\rangle$. As to eq's (\ref{rinvhom}) they
present no difficulties and their solution can be written
immediately as $h_3=h_3(I_n)$ where $I_n$ is a complete set of
SO($d$) group invariants which can be constructed by using ${\bf
P}^I$ and $\alpha^I$. A real difficulty is to choose then  such
$h_3(I_n)$ which satisfies locality condition (\ref{loc}). We
succeeded  in finding such the solution by breaking manifest
SO($d$) invariance  and keeping manifest symmetry only for an
SO($d-2$) subgroup. Due to lack of space we cannot go into much
detail concerning motivations which lead to specific form of
solution below.  Instead we first present result and then
demonstrate how this result works in particular cases.

First, we decompose ${\bf P}^I$ according to
the rule
${\bf P}^I\rightarrow \{
{\bf P}^{z},\,{\bf P}^{\bar{z}},\, {\bf P}^i\}$,
$i=3,\ldots, d$.
Then we use the well known fact that each representation of
SO($d$) group can be realized as induced representation by
inducing from SO($d-2$)$\otimes$SO(2) subgroup. The
representation for generators $M^{IJ}$ obtained in such a way
is given by
\begin{eqnarray}
&&{\rm i}M^{zi}=\bar{\zeta}^i\,,
\qquad\qquad\quad
{\rm i}M^{\bar{z}i}
=
-\frac{1}{2}\zeta^2 \bar{\zeta}^i
+\zeta^i (\zeta\bar{\zeta})
+{\rm i}S^{ij}\zeta^j
-\lambda\zeta^i\,,
\nonumber\\
\label{indgen} \\
[-7pt]
&&{\rm i}M^{z\bar{z}}=-(\zeta\bar{\zeta})+\lambda\,,
\qquad
{\rm i}M^{ij}
=\zeta^i \bar{\zeta}^j
-\zeta^j \bar{\zeta}^i+{\rm i}S^{ij}\,,
\nonumber
\end{eqnarray}
where the $S^{ij}$ stand for generators of SO($d-2$)
group, while $\lambda>0$ stands for eigevalue of SO(2)
group generator.
Note that if $M^{IJ}$ are generators in SO($d$) group
irreps labeled by ${\bf m}=(m_1,\ldots, m_\nu)$ then $S^{ij}$
are generators in SO($d-2$) group irreps labeled by
$\bar{{\bf m}}=(m_2,\ldots, m_\nu)$ for odd $d$ and
$\bar{{\bf m}}=(m_2,\ldots,m_{\nu-1}, -m_\nu)$ for even $d$.
Besides it turns out that $\lambda$ is equal to $m_1$.
The creation and
annihilation operators $\zeta^i$ and $\bar{\zeta}^i$ satisfy the
commutation relation $[\bar{\zeta}^i,\,\zeta^j] =\delta^{ij}$.
Now the solution we found can be read as follows
\begin{equation}\label{comsol1}
|h_3\rangle=({\bf P}^{\bar{z}})^k
\sum_{n=0}^{k}
\rho^n
\frac{\Gamma(\frac{D}{2}+k-n-2)}{2^n n!\Gamma(\frac{D}{2}+k-2)}
({\bf M}^{\bar{z}j}{\bf M}^{\bar{z}j})^n
V_0|0\rangle_1|0\rangle_2|0\rangle_3\,,
\end{equation}
$$
V_0\equiv
\exp(-{\rm i}q^j {\bf M}^{\bar{z}j})\tilde{V}_0\,,
\qquad
\tilde{V}_0\equiv\prod_a\beta_a^{-{\rm
i}M_a^{z\bar{z}}}G({\bf X},S)\,,
$$
where the operators ${\bf M}^{IJ}$
defined by ${\bf M}^{IJ}=\sum_{a=1}^3 M_a^{IJ}$, while
$$
{\bf X}^i\equiv\sum_a \zeta_a^i\,,
\qquad
q^i\equiv\frac{{\bf P}^i}{{\bf P}^{\bar{z}}}\,,
\qquad
\rho\equiv\frac{{\bf P}^I{\bf P}^I}{2({\bf P}^{\bar{z}})^2}\,.
\eqno{}$$
For explanation why we introduce $\tilde{V}_0$ see below.
In the solution above the $G({\bf X},S)$ is a invariant
of SO($d-2$) group transformations,
\begin{equation}\label{rotinv3}
({\rm i}{\bf X}^j\frac{\partial}{\partial {\bf X}^i}-
{\rm i}{\bf X}^i\frac{\partial}{\partial {\bf X}^j}
+\sum_a S_a^{ij})G({\bf X},S)=0\,,
\end{equation}
subjected to the following homogeneity condition
\begin{equation}\label{homcon3}
({\bf X}^j\frac{\partial}{\partial {\bf X}^j}+k-\sum_a
\lambda_a)G({\bf X},S)=0\,.
\end{equation}

Before we proceed let us comment on solution
obtained. One of interesting properties of the
solution is that the final expressions have manifest SO($d-2$)
symmetry. Another interesting property is related to
$\tilde{V}_0$.  It turns out that the $\tilde{V}_0$ has some
features of higher weight vector. Namely it satisfies the
following equations
$$
({\bf M}^{z\bar{z}}+{\rm i}k)\tilde{V}_0=0\,,
\qquad
{\bf M}^{ij}\tilde{V}_0=0\,,
\qquad
{\bf M}^{zi}\tilde{V}_0=0\,.
$$
The 1st and 2nd equations tell us that $\tilde{V}_0$
is a eigenvector of ${\bf M}^{z\bar{z}}$ and
${\bf M}^{ij}$, while the 3rd equation tells
that $\tilde{V}_0$ is a highest vector with respect to
${\bf M}^{zi}$.
Note that ${\bf M}^{\bar{z}i}\tilde{V}_0={\kern-2.3ex/}\,0$
and
$({\bf M}^{z\bar{z}}
+{\rm i}(k-1)){\bf M}^{\bar{z}i}\tilde{V}_0=0$.
The expression (\ref{comsol1}) by itself provides complete
solution to cubic interaction vertices for all higher as well
lower massless spin fields.  Making use of particular
realizations of representation of SO($d$) group and
(\ref{comsol1}) one can derive  various representation for cubic
vertices. Let us demonstrate how our solution works for the
cases of five and six dimensional space--times.

$D=5$ $case$. For this case the indices $i,j$ which
label $d-2$ directions  take one value $i,j=3$.
To simplify our
expressions we use a notation ${\bf X}={\bf X}^3$,
$\zeta=\zeta^3$ For the case under consideration the equation
(\ref{rotinv3}) does not impose of any restrictions while from
equation (\ref{homcon3}) we learn $G({\bf X})={\bf X}^{J-k}$,
where $J\equiv\sum_a \lambda_a$\,.  Now the relevant generators
(see (\ref{indgen})) read as follows
\begin{equation}
{\rm i}M^{\bar{z}3}
=\frac{1}{2}\zeta^2\bar{\zeta}-\lambda\zeta\,,
\qquad
{\rm i} M^{z\bar{z}}=-\zeta\bar{\zeta}+\lambda\,.
\end{equation}
Making use of easily derived relations
\begin{equation}\label{rel1}
\!\! \prod_a \beta_a^{-{\rm i} M_a^{z\bar{z}}}f({\bf X})
=\prod_a\beta_a^{-\lambda_a}f({\bf X}(\beta))\,,
\quad
\!\!e^{-q(\zeta^2\bar{\zeta}+b\zeta)}f(\zeta)
=(1+q\zeta)^{-b}f\Bigl(\frac{\zeta}{1+q\zeta}\Bigr)\,,
\end{equation}
where ${\bf X}(\beta)=\sum_a \beta_a\zeta_a$ we get
\begin{equation}\label{civ5}
V_0
=\prod_a \beta_a^{-\lambda_a}(1+q\zeta_a)^{2\lambda_a}
\Bigr(\sum_a\frac{\beta_a\zeta_a}{1+q\zeta_a}\Bigl)^{J-k}\,,
\end{equation}
where $q\equiv q^1/2$. The expressions (\ref{civ5})
and (\ref{comsol1}) provide complete set of cubic interaction
vertices for the massless higher spin fields living in a
flat $5D$ space--time. Now we move to $6D$ space--time.

$D=6$ $case$.
For this case the indices $i,j$
take two values. We prefer to pass to complex coordinates and now
indices $i,j$ run over $x$ and $\bar{x}$.
The generator of the SO(2) group can be presented as
$S^{x\bar{x}}={\rm i}\tau$.
Note that $\lambda$ and $\tau$ are linked with highest weight
vector ${\bf m}=(m_1,m_2)$ as follows $\lambda=m_1$,
$\tau=-m_2$.
Now ${\bf X}^i=\{{\bf X}^x,\,{\bf X}^{\bar{x}}\}$ and we use the
following notation
$$
j_{a1}\equiv (\lambda_a-\tau_a)/2\,,\quad
j_{a2}\equiv(\lambda_a+\tau_a)/2\,,\quad
{\bf J}_\sigma\equiv\sum_a j_{a\sigma}\,,\quad
J_a\equiv\sum_{\sigma=1,2} j_{a\sigma}\,,
\eqno{}$$
and $q_1\equiv q^x$,
$q_2\equiv q^{\bar{x}}$,
$\zeta_{a1}\equiv \zeta^{\bar{x}}_a$,
$\zeta_{a2}\equiv\zeta^x_a$.
Solution to equations (\ref{rotinv3},\ref{homcon3}) is
$$
G({\bf X},\tau)=({\bf X}^{\bar{x}})^{{\bf J}_1-\frac{k}{2}}
({\bf X}^x)^{{\bf J}_2-\frac{k}{2}}\,.
\eqno{}$$
The relevant generators are
$$
{\rm i} M^{\bar{z}\bar{x}}
=(\zeta^{\bar{x}})^2\bar{\zeta}^x
-(\lambda-\tau)\zeta^{\bar{x}}\,,
\qquad
{\rm i} M^{\bar{z}x}=(\zeta^x)^2\bar{\zeta}^{\bar{x}}
-(\lambda+\tau)\zeta^x\,.
$$
Now making use of the relations (\ref{rel1})
we get
\begin{equation}\label{civ6}
V_0
=\prod_a \beta_a^{-J_a}\prod_{\sigma a}
(1+q_\sigma \zeta_{a \sigma})^{2j_{a\sigma}}
\prod_\sigma\Bigr(\sum_a
\frac{\beta_a\zeta_{a\sigma}}{1
+q_\sigma\zeta_{a\sigma}}\Bigl)^{{\bf J}_\sigma-\frac{k}{2}}\,.
\end{equation}

Note that while our solution has SO($d$) symmetry it keeps
manifest symmetry only for SO($d-2$) group.
Besides as it is seen from (\ref{comsol1}) there is manifest
locality with respect to transversal momenta ${\bf P}^i$ and
${\bf P}^z$. It turns out that for the vertices to be
local with respect
${\bf P}^{\bar{z}}$  it is necessary to impose certain
restrictions on $k$ and spin value $j$. By direct inspection of
expressions (\ref{civ5}) and (\ref{civ6}) one can make sure
that the restrictions in questions are
\begin{equation}\label{restrs}
D=5:\qquad
J-2\lambda_a\leq k \leq J \,;\qquad
D=6:\qquad
2({\bf J}_\sigma-2j_{a\sigma})\leq k \leq 2{\bf J}_\sigma\,.
\end{equation}

In conclusion, let us note that the inequalities
(\ref{restrs})  for $k=2$ are precisely the
restrictions on $j$  which leave no place for the
gravitation interaction of massless higher spin
($\lambda,j>2$) fields. In spite of this there exists a wide
class of Pauli like cubic interaction vertices allowed by
inequalities (\ref{restrs}) and, we believe, the theory under
consideration can get interesting applications and deserves a
further investigation.

{\it Acknowledgements}.
This work was supported in part by the European Community
Commission under the contract INTAS, Grant No.94-2317, by the
Russian Foundation for Basic Research, Grant No.96-02-17314-a,
and by the NATO Linkage, Grant No.931717.

\end{document}